\documentclass{SciPost}

\hypersetup{
    colorlinks,
    linkcolor={red!50!black},
    citecolor={blue!50!black},
    urlcolor={blue!80!black}
}

\usepackage{braket}
\usepackage[bitstream-charter]{mathdesign}
\DeclareSymbolFont{usualmathcal}{OMS}{cmsy}{m}{n}
\DeclareSymbolFontAlphabet{\mathcal}{usualmathcal}

\fancypagestyle{SPstyle}{
\fancyhf{}
\lhead{\colorbox{scipostblue}{\bf \color{white} ~SciPost Physics }}
\rhead{{\bf \color{scipostdeepblue} ~Submission }}

\fancyfoot[C]{\textbf{\thepage}}
}

\begin{document}

\pagestyle{SPstyle}

\begin{center}{\Large \textbf{\color{scipostdeepblue}{
Enhanced squeezing for quantum gravimetry in a Bose-Einstein condensate with focussing\\
}}}\end{center}

\begin{center}\textbf{
Lewis A. Williamson\textsuperscript{1},
Karandeep Gill\textsuperscript{2$\star$},
Andrew J. Groszek\textsuperscript{1},
Matthew J. Davis\textsuperscript{1} and
Simon Haine\textsuperscript{2$\dagger$}
}\end{center}

\newcommand\blfootnote[1]{
  \begingroup
  \renewcommand\thefootnote{}\footnote{#1}
  \addtocounter{footnote}{-1}
  \endgroup
}

\blfootnote{\textsuperscript{$\star$} Current address: Q-CTRL, 93 Shepherd St, Chippendale, NSW 2008, Australia}

\begin{center}
{\bf 1} ARC Centre of Excellence for Engineered Quantum Systems, School of Mathematics and Physics, University of Queensland, St Lucia, Queensland 4072, Australia
\\
{\bf 2} Department of Quantum Science and Technology, Research School of Physics, Australian National University
\\[\baselineskip]
$\dagger$ \href{mailto:email1}{\small simon.haine@anu.edu.au}
\end{center}

\section*{\color{scipostdeepblue}{Abstract}}
\textbf{\boldmath{
Free-fall atom interferometers offer a powerful platform for accurate, absolute gravitational sensing. Szigeti \emph{et al}.~[Phys.\ Rev.\ Lett.\ \textbf{125}, 100402 (2020)] recently proposed a quantum-enhanced scheme that uses a spin-squeezed Bose-Einstein condensate as an input state to improve the phase sensitivity of the interferometer. The spin squeezing, generated via one-axis twisting interactions, was limited by condensate expansion. Here we present an improved state preparation in which a sudden trapping potential---a delta kick---is initially applied to focus the condensate. The resulting increase in density enhances the one-axis-twisting interactions and produces greater spin squeezing. Using multimode truncated-Wigner simulations, we quantify the performance of the interferometer and find that, for an optimal kick strength, the phase sensitivity surpasses the standard quantum limit by a factor of $\sim 20$. This represents a fourfold improvement over the original scheme without the delta kick and is well captured by a two-mode approximation.
}}

\vspace{\baselineskip}

\noindent\textcolor{white!90!black}{%
\fbox{\parbox{0.975\linewidth}{%
\textcolor{white!40!black}{\begin{tabular}{lr}%
  \begin{minipage}{0.6\textwidth}%
    {\small Copyright attribution to authors. \newline
    This work is a submission to SciPost Physics. \newline
    License information to appear upon publication. \newline
    Publication information to appear upon publication.}
  \end{minipage} & \begin{minipage}{0.4\textwidth}
    {\small Received Date \newline Accepted Date \newline Published Date}%
  \end{minipage}
\end{tabular}}
}}
}


\vspace{10pt}
\noindent\rule{\textwidth}{1pt}
\tableofcontents
\noindent\rule{\textwidth}{1pt}
\vspace{10pt}


\section{Introduction} Free-fall atom interferometers can provide incredibly sensitive measurements of inertial quantities, making them well-suited for gravimetry~\cite{Peters:1999,Peters:2001,Hu:2013,Hauth:2013,Altin:2013,Farah:2014,Freier:2016,Menoret:2018,Zhang:2023,Saywell:2023}, gravitational and magnetic gradiometry~\cite{Snadden:1998,Sorrentino:2014,Biedermann:2015,Hardman:2016b,DAmico:2016,Asenbaum:2017,Janvier:2022}, accelerometry~\cite{Canuel:2006,Templier:2022}, and rotation sensing \cite{Gustavson:1997,Gustavson:2000,Durfee:2006,Gauguet:2009,Barrett:2014,Gautier:2022,Woffinden:2023}. Advancements in robustness, portability and autonomous operation would enable new capabilities in a range of applications~\cite{Bongs:2019}, including navigation~\cite{Jekeli:2005,Battelier:2016,Narducci:2022,Wright:2022,Phillips:2022, Wang:2023}, mineral exploration~\cite{Van:2000,Evstifeev:2017}, and hydrology~\cite{Canuel:2018,Schilling:2020}. In laboratory applications, atom-based inertial sensors have been used in tests of the equivalence principal~\cite{Aguilera:2014,Williams:2016,Becker:2018} and measurements of the gravitational constant~\cite{Rosi:2014}, while tests of general relativity would be made possible with improved sensitivity~\cite{Dimopoulos:2007,Tino:2021}.

Standard atom interferometers use uncorrelated atoms, and hence their precision is bounded by the shot noise limit. Quantum entanglement offers a promising pathway to improved measurement sensitivity, as it allows the shot noise limit to be surpassed~\cite{Wineland:1992, Pezze_review:2018, Giovannetti:2004, Szigeti:2021, Corgier:2023, Corgier:2025}. Bose–Einstein condensates (BECs) are an excellent candidate for entanglement‑enhanced interferometry, due to their high level of control and tunability~\cite{Pezze_review:2018}. Indeed, numerous experiments have demonstrated the generation of entanglement and spin squeezing in BECs and ensembles of cold atoms, often yielding clear interferometric enhancement, but without implementing a complete metrological measurement that exploits this enhancement in a practical sensing task~\cite{Appel:2009,Leroux:2010,Riedel:2010,Gross:2010,Strobel:2014}. In practice, extending such techniques to free‑fall atom interferometers remains challenging, as the entanglement generation must be compatible with the coherent splitting of the atomic ensemble into distinct momentum modes~\cite{Anders:2021}. Experiments with BECs of entangled atoms have demonstrated sensitivity below the standard quantum limit in magnetometry~\cite{Ockeloen:2013,Muessel:2014}, atomic clocks~\cite{Kruse:2016} and, very recently, in gravimetry~\cite{Cassens:2025}.

\begin{figure}
\includegraphics[trim=0cm 0cm 0cm 0cm,clip=true,width=\textwidth]{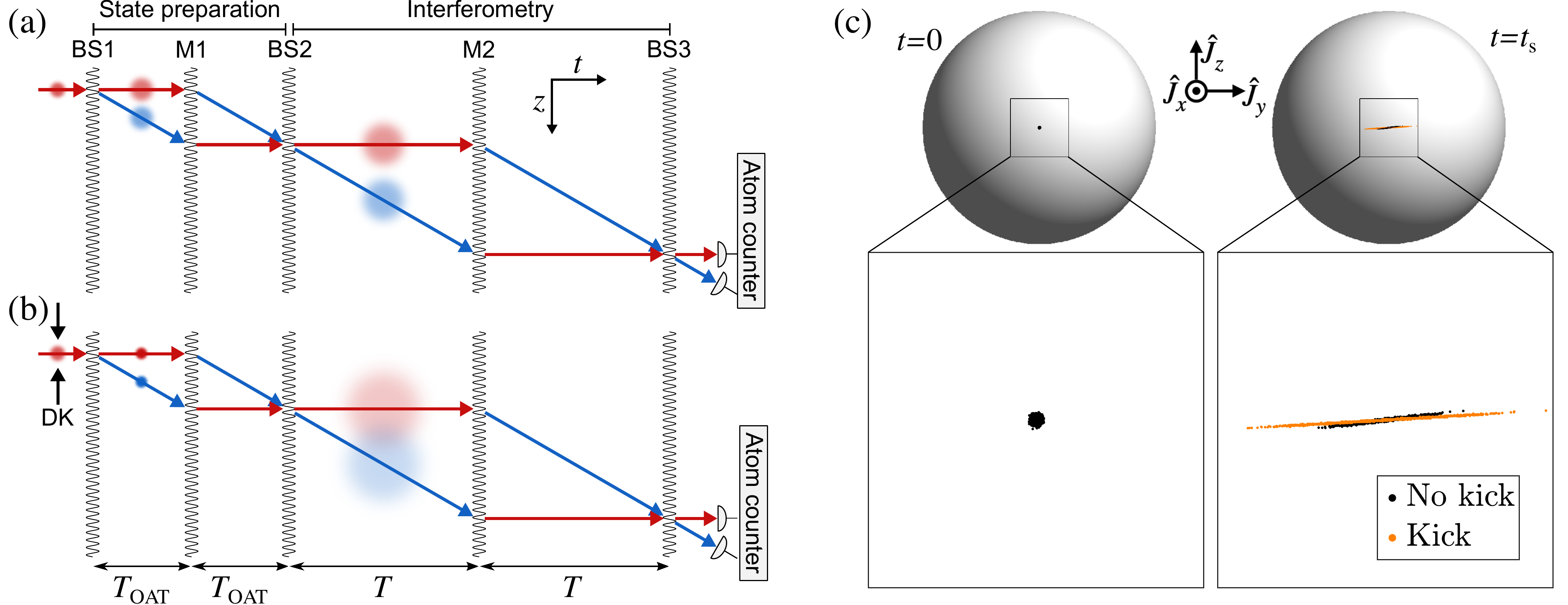}
\caption{\label{fig:scheme} (a) In the scheme of Szigeti \emph{et al}.~\cite{Szigeti:2020}, OAT interactions in a BEC are used during state preparation to produce a spin-squeezed state with enhanced phase sensitivity. The protocol then continues in a standard Mach-Zehnder interferometry scheme. The squeezing is limited by condensate expansion during state preparation, which reduces OAT interactions. (b) In this paper we modify the state preparation by including an initial delta kick (DK), which causes the component densities to initially focus. This increases OAT interactions, resulting in increased spin squeezing and phase sensitivity. (c) Scatter plot of $(\mathcal{J}_x,\mathcal{J}_y,\mathcal{J}_z)$ on the surface of the Bloch sphere at times $t=0$ and $t=t_\mathrm{s}\approx 2T_\mathrm{OAT}$, obtained from $10^4$ truncated Wigner trajectories. The distribution is clearly squeezed at the state preparation end time $t_\mathrm{s}$, which is enhanced by an initial delta kick. Square frames show the $(\mathcal{J}_y,\mathcal{J}_z)$ distribution magnified; for simplicity we remove equatorial drift and hence the distributions are centred at $\langle\hat{J}_y\rangle=0$.}
\end{figure}

Szigeti \emph{et al}.~\cite{Szigeti:2020} have proposed a quantum-enhanced gravimetry scheme, shown schematically in Fig.~\ref{fig:scheme}(a), that uses spin-squeezed states in a BEC, generated via the one-axis twisting (OAT) mechanism~\cite{Kitagawa:1993}. The resulting sensitivity was between two and five times below the standard quantum limit, depending on the atom number. The spin squeezing, however, was limited by the free expansion of the BEC, which decreases the OAT interactions. To mitigate this, Corgier \emph{et al}.~\cite{Corgier:2021} proposed focussing the condensate with a rapidly applied trapping potential --- a delta kick~\cite{Ammann:1997,Muntinga:2013,Kovachy:2015,Deppner:2021,Anders:2021,Cassens:2025} --- which increases the density and therefore the OAT interactions. Using the delta kick to increase the density, rather than relying on a dense initial state, means that the density can be low while the clouds are separating, reducing deleterious intercomponent scattering during the initial beamsplitter process. The scheme in~\cite{Corgier:2021} achieved a high theoretical sensitivity, but applied a delta kick to each arm of the interferometer, after the initial beamsplitter and separation of the atoms. As the delta kick is applied after a relative phase has been established, slight variations or asymmetries in the strength of the delta-kick potential applied to each arm could reduce mode matching and result in variable phase-shifts that wash out the squeezing.

In this paper we introduce a delta-kick scheme that overcomes the practical limitations of~\cite{Corgier:2021}. In this modified scheme, shown schematically in Fig.~\ref{fig:scheme}(b), the delta kick is applied \emph{before} the initial beamsplitting process while the atoms are harmonically confined. Centering on the atomic cloud is therefore straightforward and, most importantly, the effect of the kick is symmetric across the two interferometer arms. We show that an optimal kick strength improves phase sensitivity by a factor of $\sim 20$ compared to the standard quantum limit --- a fourfold improvement on the scheme with no kick~\cite{Szigeti:2020} and within a factor of two of~\cite{Corgier:2021}. Furthermore, a delta kick reduces the time needed to generate spin squeezing, and hence a shorter state-preparation time is required. Our results are obtained using truncated Wigner simulations, which account for multi-mode dynamics, imperfect mode overlap, and phase diffusion. At the optimal kick strength the spin squeezing is close to that obtained from a two-mode approximation.

This paper is organised as follows. In Sec.~\ref{sec:scheme} we introduce the standard Mach-Zehnder based gravimetry scheme and discuss how this can be enhanced using spin squeezing. We then introduce the delta-kick scheme that will be the subject of this paper. Section~\ref{sec:numerics} presents the truncated Wigner method and further numerical details. Our results are presented in Sec.~\ref{sec:results}, where we characterise the effect of the delta kick and show how this can enhance spin squeezing. We compare the truncated Wigner simulations with a two-mode approximation and discuss effects of varying the initial trap frequency and the state-preparation time. We conclude in Sec.~\ref{sec:conclusion}.

\section{Quantum-enhanced gravimetry and the delta-kick scheme}\label{sec:scheme}

\subsection{Gravimetry using a Mach-Zehnder atom interferometer}
A Mach-Zehnder atom interferometer uses the interference of two hyperfine states $\ket{1}$ and $\ket{2}$ in an atomic gas to measure gravitational acceleration~\cite{Kasevich:1992}. The interferometer consists of a $\pi/2$--$\pi$--$\pi/2$ (beamsplitter--mirror--beamsplitter) pulse sequence. The pulses are implemented using counter-propagating lasers (wavevectors $\mathbf{k}_1$ and $\mathbf{k}_2$ respectively), which coherently couple the two states $\ket{1}$ and $\ket{2}$ via two-photon Raman transitions. Crucially, this transfers momentum $\hbar(\mathbf{k}_1-\mathbf{k}_2)\equiv \hbar k_0\hat{\mathbf{z}}$ to the atoms in component $\ket{2}$ (for simplicity we assume the transferred momentum is in the free-fall direction $\hat{\mathbf{z}}$). During free fall, the two interferometry arms therefore acquire a relative phase $\phi=k_0gT^2$, with $g$ the gravitational acceleration and $2T$ the total time of the interferometry sequence. This relative phase can be inferred from a number difference measurement at the interferometer output, and the gravitational acceleration can then be determined with a sensitivity of $\Delta\phi/(k_0 T^2)$, with $\Delta\phi$ the uncertainty in the relative phase~\cite{Kasevich:1992}. In the standard scheme~\cite{Kasevich:1992} the atoms undergo expansion prior to interferometry to reduce interactions and correlations. The uncertainty in the final relative phase measurement is then given by the standard quantum limit
\begin{equation}\label{eq:sql}
    \Delta\phi=\Delta \phi_\mathrm{SQL} \equiv \frac{1}{\sqrt{N}},
\end{equation}
with $N$ the number of atoms.

\subsection{Quantum-enhanced gravimetry}\label{sec:QEG}
In the scheme of Szigeti \emph{et al}.~\cite{Szigeti:2020}, the atomic gas is Bose condensed and the initial free-expansion stage is replaced with a more involved state-preparation [see Fig.~\ref{fig:scheme}(a)]. Atomic interactions, which in the standard scheme are deliberately removed, are instead exploited to generate quantum correlations between the two arms of the interferometer via OAT interactions~\cite{Kitagawa:1993,Sorensen:2001b}. The resulting spin-squeezed state leads to reduced number-difference fluctuations at the output of the interferometer, and therefore a more precise phase estimate. We outline the details of the scheme of~\cite{Szigeti:2020} below, before describing the improvements made in this work.

\subsubsection{State preparation}
The state preparation begins with all atoms Bose condensed in the first component $\ket{1}$ and confined by a harmonic trap. The atoms are described by a spinor field
\begin{equation}
    \hat{\Psi}(\textbf{r},t) = \left(\begin{array}{c}\hat{\psi}_1 (\textbf{r},t)\\\hat{\psi}_2 (\textbf{r},t)\end{array}\right),
\end{equation}
with field operators $\hat{\psi}_j$ describing the two atomic states $\ket{j}$ and defined in the free-falling frame. The condensate is then released from the harmonic trap and a $\pi/2$ pulse is applied via the unitary $\hat{V}_{\frac{\pi}{2},-\frac{\pi}{2}}$ (BS1 in Fig.~\ref{fig:scheme}), with
\begin{subequations}\label{eq:rotate}
\begin{align}
    \hat{V}_\mathrm{\theta,\phi}^\dagger\hat{\psi}_1\hat{V}_\mathrm{\theta,\phi}&=\cos\left(\frac{\theta}{2}\right)\hat{\psi}_1-ie^{i\phi}\sin\left(\frac{\theta}{2}\right)e^{-ik_0z}\hat{\psi}_2,\\
    \hat{V}_\mathrm{\theta,\phi}^\dagger\hat{\psi}_2\hat{V}_\mathrm{\theta,\phi}&=\cos\left(\frac{\theta}{2}\right)\hat{\psi}_2-ie^{-i\phi}\sin\left(\frac{\theta}{2}\right)e^{ik_0z}\hat{\psi}_1.
\end{align}
\end{subequations}
This produces an equal superposition of the two hyperfine states.

The condensate is then evolved in time, during which atomic interactions produce OAT. First, the two components are evolved until a time $T_\mathrm{OAT}$ via the unitary $\hat{U}_{0,T_\mathrm{OAT}}$, with $\hat{U}_{t,t^\prime}$ the time evolution operator from time $t$ to $t^\prime$. During this time the two components spatially separate due to their relative momentum $\hbar k_0\hat{\mathbf{z}}$ imparted by BS1. To recombine the two components, a $\pi$ pulse is applied at $t=T_\mathrm{OAT}$ via the unitary $\hat{V}_{\pi,0}$ (M1 in Fig.~\ref{fig:scheme}) and the state is evolved until a time $t_\mathrm{s}\approx 2T_\mathrm{OAT}$ via the unitary $\hat{U}_{T_\mathrm{OAT},t_\mathrm{s}}$ (the precise time $t_\mathrm{s}$ is chosen to optimise the final measurement sensitivity). When the components recombine, the spin distribution is squeezed due to the accumulated OAT, and the resulting spin-squeezed state can be used for interferometry. To summarise, the field operator throughout the state preparation is (suppressing spatial dependence)
\begin{equation}\label{eq:timeevolve}
\hat{\Psi}(t)=\left\{\begin{array}{ll}\hat{U}_{0,t}^\dagger\hat{V}_{\frac{\pi}{2},-\frac{\pi}{2}}^\dagger\hat{\Psi}(0)\hat{V}_{\frac{\pi}{2},-\frac{\pi}{2}}\hat{U}_{0,t},&0<t\le T_\mathrm{OAT},\\
\\
\hat{U}_{T_\mathrm{OAT},t}^\dagger\hat{V}_{\pi,0}^\dagger\hat{\Psi}(T_\mathrm{OAT})\hat{V}_{\pi,0}\hat{U}_{T_\mathrm{OAT},t},&T_\mathrm{OAT}<t\le t_\mathrm{s}.\end{array}\right.
\end{equation}

Spin squeezing manifests in the variances of the pseudo-spin operators $\hat{J}_\nu$ obtained from the time-evolved field Eq.~\eqref{eq:timeevolve},
\begin{subequations}\label{eq:spinops}
\begin{align}
    \hat{J}_x(t) &= \frac{1}{2}\int\left(\hat{\psi}_1^\dagger(\mathbf{r},t)\hat{\psi}_2(\mathbf{r},t)e^{-ik_0z}+\hat{\psi}_2^\dagger(\mathbf{r},t)\hat{\psi}_1(\mathbf{r},t)e^{ik_0z}\right)d\textbf{r},\\
    \hat{J}_y(t) &=-\frac{i}{2}\int\left(\hat{\psi}_1^\dagger(\mathbf{r},t)\hat{\psi}_2(\mathbf{r},t)e^{-ik_0z}-\hat{\psi}_2^\dagger(\mathbf{r},t)\hat{\psi}_1(\mathbf{r},t)e^{ik_0z}\right)d\textbf{r},\\
    \hat{J}_z(t) &= \frac{1}{2}\int\left(\hat{\psi}_1^\dagger(\mathbf{r},t)\hat{\psi}_1(\mathbf{r},t)-\hat{\psi}_2^\dagger(\mathbf{r},t)\hat{\psi}_2(\mathbf{r},t)\right)d\textbf{r}.
\end{align}
\end{subequations}
An example of the spin dynamics during the state preparation is shown in Fig.~\ref{fig:scheme}(c). Each point in the two scatter plots corresponds to an individual trajectory obtained from sampling the Wigner function (see Sec.~\ref{sec:TW}). The distribution is squeezed at $t=t_\mathrm{s}$, evidenced by the strongly elliptical shape, with the squeezing enhanced by a delta kick.

\subsubsection{Interferometry sequence}
The state preparation is followed by a beamsplitter--mirror--beamsplitter interferometry sequence of duration $2T$. This begins with a rotation at the second beamsplitter (BS2 in Fig.~\ref{fig:scheme}). The components then separate for a time $T$, before a second $\pi$ pulse is applied (M2 in Fig.~\ref{fig:scheme}), causing the components to recombine at time $2T$ at the final beamsplitter (BS3 in Fig.~\ref{fig:scheme}). The relative phase accumulated during the interferometry sequence is determined from a number-difference measurement at the atom counter, with the angles of the two beamsplitters BS2 and BS3 chosen to maximise the measurement sensitivity.

\subsubsection{Measurement sensitivity}
We assume that the condensate is sufficiently dilute during the interferometry sequence that atomic interactions are negligible. In the limit $T\gg T_\mathrm{OAT}$, the optimised measurement sensitivity at the interferometer output is then directly proportional to the spin squeezing accumulated during the state preparation~\cite{Szigeti:2020},
\begin{equation}\label{eq:phase}
    \Delta \phi =\frac{\xi}{\sqrt{N}}=\xi\Delta\phi_\mathrm{SQL},
\end{equation}
with $\Delta\phi_\mathrm{SQL}$ the sensitivity of uncorrelated atoms [Eq.~\eqref{eq:sql}]. Here
\begin{equation}
\xi = \operatorname{min}_{\theta,\phi,t}\frac{\sqrt{N\mathrm{Var}(\hat{J}_{\theta,\phi}(t))}}{\left|\langle \hat{J}_{\frac{\pi}{2},\phi+\frac{\pi}{2}}(t)\rangle\right|}\equiv\frac{\sqrt{N\mathrm{Var}(\hat{J}_\perp(t_\mathrm{s}))}}{\left|\langle \hat{J}_\parallel(t_\mathrm{s})\rangle\right|}\label{xi1}
\end{equation}
is the Wineland spin-squeezing parameter~\cite{Wineland:1992,Wineland:1994,Sorensen:2001b} minimised over Bloch sphere angles $\theta,\phi$ and time $t$, with $\operatorname{Var}(\hat{X})=\langle\hat{X}^2\rangle-\langle\hat{X}\rangle^2$ and
\begin{equation}
\hat{J}_{\theta,\phi}(t)=\sin\theta\sin\phi\hat{J}_x(t)+\sin\theta\cos\phi\hat{J}_y(t)+\cos\theta\hat{J}_z(t).
\end{equation}
The minimisation over time in Eq.~\eqref{xi1} selects the optimal time $t_\mathrm{s}$ to end the state preparation, and depends on the strength of the delta kick. For notational convenience, we have defined $\langle\hat{J}_\perp\rangle\equiv\langle\hat{J}_{\theta_\mathrm{s},\phi_\mathrm{s}}\rangle$ and $\langle\hat{J}_\parallel\rangle\equiv\langle\hat{J}_{\frac{\pi}{2},\phi_\mathrm{s}+\frac{\pi}{2}}\rangle$, with $\theta_\mathrm{s}$ and $\phi_\mathrm{s}$ obtained from the minimisation over angles in Eq.~\eqref{xi1}. These angles describe the orientation and position of the squeezed elliptical distribution at the end of the state preparation [see Fig.~\ref{fig:scheme}(c)], with $\theta_\mathrm{s}$ the angle of the minor axis (the direction of reduced variance) and $\phi_\mathrm{s}$ the equatorial drift of the spin vector. By appropriately choosing the angles of the second (BS2) and third (BS3) beamsplitters in Fig.~\ref{fig:scheme}, the spin-squeezed state can be rotated so that it is maximally sensitive to a number difference measurement at the interferometer output, resulting in the sensitivity Eq.~\eqref{eq:phase}.

\subsubsection{Limitations and delta-kick focussing scheme}
The protocol of~\cite{Szigeti:2020} was shown to increase phase sensitivity five-fold when compared to the standard interferometry scheme. However, the amount of OAT was limited by the  expansion of the atomic cloud, which reduces the strength of interactions. The delta-kick scheme we introduce here employs an initial rapid tightening of the harmonic trap, prior to releasing the condensate, see Fig.~\ref{fig:scheme}(b). This gives the cloud inward momentum, causing it to initially contract and become denser, before expansion due to interactions dominates at later times. The higher atomic density increases the OAT interactions, resulting in enhanced spin squeezing, see Fig.~\ref{fig:scheme}(c). Apart from this modification, the scheme is identical to that in~\cite{Szigeti:2020} described above.

\section{Numerical details}\label{sec:numerics}

\subsection{Truncated Wigner method}\label{sec:TW}
We simulate the state preparation scheme using the truncated Wigner method~\cite{Steel:1998, Sinatra:2002, Blakie:2008}. This enables calculation of the squeezing parameter, Eq.~\eqref{xi1}, from which we can determine the measurement sensitivity [Eq.~\eqref{eq:phase}] \cite{Haine:2011, Haine:2018}. An ensemble of initial fields $\psi_j(\mathbf{r},0)$, which incorporate both a coherent part and quantum fluctuations, are evolved using the Gross-Pitaevskii equation (GPE) in the free-falling frame. According to the truncated Wigner prescription, we have~\cite{Blakie:2008}
\begin{equation}\label{eq:TWdensities}
    \langle \mathcal{C}[\hat{\psi}_1,\hat{\psi}_1^\dagger,\hat{\psi}_2,\hat{\psi}_2^\dagger]\rangle_\mathrm{sym}=\overline{\mathcal{C}[\psi_1,\psi_1^*,\psi_2,\psi_2^*]},
\end{equation}
where $\mathcal{C}$ is a field correlation, ``sym'' denotes symmetric ordering of operators (for example $\langle\hat{\psi}_1^\dagger\hat{\psi}_1\rangle_\mathrm{sym}=\frac{1}{2}(\langle\hat{\psi}_1^\dagger\hat{\psi}_1\rangle+\langle\hat{\psi}_1\hat{\psi}_1^\dagger\rangle)$, and an overbar denotes an average over truncated Wigner trajectories. Equation~\eqref{eq:TWdensities} gives
\begin{subequations}\label{eq:TWexpect}
    \begin{align}
        \langle\hat{J}_\nu\rangle&=\overline{\mathcal{J}_\nu},\\
        \langle\hat{J}_\nu\hat{J}_\kappa\rangle&=\overline{\mathcal{J}_\nu \mathcal{J}_\kappa}-\frac{m}{8}\delta_{\nu\kappa},
    \end{align}
\end{subequations}
where $\mathcal{J}_\nu$ is $\hat{J}_\nu$ [Eq.~\eqref{eq:spinops}] but with $\hat{\psi}_j\rightarrow\psi_j$, and $m$ is the total number of numerical modes. The spin squeezing Eq.~\eqref{xi1} is then calculated from the expectation values Eq.~\eqref{eq:TWexpect}.

The validity of truncated Wigner requires many more particles than modes, i.e.\ $N\gg m$~\cite{Sinatra:2002}. This makes three-dimensional simulations problematic, as the large phase-space volume requires a large number of numerical modes, which gives $m>N$ for reasonable condensate sizes~\cite{Ferris:2008}. We instead use an effective one-dimensional description based on~\cite{Castin:1996} and adapted to the current setup in~\cite{Szigeti:2020}.  First, the fields $\psi_j$ are assumed to be separable,
\begin{subequations}\label{eq:psiseparate}
\begin{align}
    \psi_1(\mathbf{r},t)&=\psi_\perp(\rho,t)\varphi_1(z,t),\\
    \psi_2(\mathbf{r},t)&=\psi_\perp(\rho,t)\varphi_2(z,t)e^{ik_0z},
\end{align}
\end{subequations}
with $z$ aligned along the free-fall direction. We assume azimuthal symmetry, such that the transverse direction is described by a single coordinate $\rho=\sqrt{x^2+y^2}$. In general, quantum noise should be added to $\psi_\perp$ and $\varphi_j$; here we make the approximation that the transverse field component $\psi_\perp$ is noiseless and all quantum fluctuations are contained in $\varphi_j$. The field $\varphi_2$ is defined in a frame with momentum $\hbar k_0\hat{\mathbf{z}}$, which is more convenient in the numerics.

The dynamics in the transverse coordinate is described by self-similar expansion starting from an initial Thomas-Fermi profile,
\begin{equation}\label{eq:TF}
    \psi_\perp(\rho,t)=\left\{\begin{array}{ll}\sqrt{\frac{2}{\pi b(\omega t)^2r_\mathrm{TF}^2}\left(1-\frac{\rho^2}{b(\omega t)^2r_\mathrm{TF}^2}\right)},&\rho\le b(\omega t)r_\mathrm{TF},\\
    \\
    0,&\rho>b(\omega t)r_\mathrm{TF}.\end{array}\right.
\end{equation}
Here $r_\mathrm{TF}=\sqrt{2\mu/M\omega^2}$ is the initial Thomas-Fermi radius ($\mu$ is the chemical potential) and the choice of normalization gives $\int |\psi_\perp(\rho,t)|^2\,dxdy=1$. The scaling factor $b(\omega t)$ satisfies
\begin{equation}\label{eq:Rdynamics}
    \ddot{b}(u)=\frac{1}{b(u)^4},
\end{equation}
with $b(0)=1$ and $\dot{b}(0)=0$. The dynamics in $z$ is then described by an effective one-dimensional GPE in the free-falling frame,
\begin{subequations}\label{eq:GPEeff}
\begin{align}
    i\hbar\frac{\partial \varphi_1(z,t)}{\partial t}&=\left[\frac{\hat{p}_z^2}{2M}+\tilde{g}_{11}(t)|\varphi_1|^2+\tilde{g}_{12}(t)|\varphi_2(z,t)|^2\right]\varphi_1(z,t),\label{GPE1}\\
    i\hbar\frac{\partial \varphi_2(z,t)}{\partial t}&=\left[\frac{(\hat{p}_z+\hbar k_0)^2}{2M}+\tilde{g}_{22}(t)|\varphi_2(z,t)|^2+\tilde{g}_{12}(t)|\varphi_1(z,t)|^2\right]\varphi_2(z,t),\label{GPE2}
\end{align}
\end{subequations}
with $\hat{p}_z=-i\hbar\partial/\partial z$ and normalization $\int (|\varphi_1(z,t)|^2+|\varphi_2(z,t)|^2)=N$. The effective interaction strengths are $\tilde{g}_{j\ell}(t)=4 g_{j\ell}/(3\pi r_\mathrm{TF}^2b(\omega t)^2)$, with $g_{j\ell} = 4\pi\hbar^2\mathrm{a}_{j\ell}/M$ the three-dimensional interaction strengths dependent on s-wave scattering lengths $\mathrm{a}_{j\ell}$.\footnote{As a result of Eq.~\eqref{eq:TWdensities}, the density terms in Eq.~\eqref{eq:GPEeff} should be modified slightly to $\tilde{g}_{jj}(|\varphi_j|^2-m/L)$ and $\tilde{g}_{12}(|\varphi_j|^2-m/2L)$. We ignore the shifts $\propto m/L$, as for our simulations they only result in a small relative phase shift $(m/L)\int_0^{t_\mathrm{s}}(\tilde{g}_{11}(\tau)-\tilde{g}_{22}(\tau))d\tau\sim 10^{-3}$ between $\varphi_1$ and $\varphi_2$ that can anyway be adjusted for by the second beamsplitter (BS2 in Fig.~\ref{fig:scheme}).}

All atoms are initially in the field $\varphi_1(z,0)$, which consists of both a coherent part $\varphi_1^\mathrm{c}(z)$ and quantum fluctuations. The coherent part $\varphi_1^\mathrm{c}(z)$ is obtained by evolving Eq.~\eqref{GPE1} in imaginary time with a trap $\frac{1}{2}M\omega^2 z^2$, $\tilde{g}_{11}=\tilde{g}_{11}(0)$, $\varphi_2=0$, and chemical potential $\mu$ tuned to obtain the desired number of atoms. Noise fields $\eta_j(z)$ are then added to account for quantum fluctuations, $\varphi_1(z,0)=\varphi_1^\mathrm{c}(z)+\eta_1(z)$ and $\varphi_2(z,0)=\eta_2(z)$. The noise fields are
\begin{equation}\label{eq:noise}
    \eta_j(z)=\frac{1}{\sqrt{2\pi}}\int \frac{p_j(k)+iq_j(k)}{2}e^{ikz}\,dz,
\end{equation}
with $p_j(k)$ and $q_j(k)$ sampled from a normal distribution with zero mean and unit variance. This choice of noise corresponds to a sampling of the free-space Wigner function and adds, on average, half a particle to each momentum mode~\cite{Blakie:2008}.

A delta kick is applied by rapidly tightening the trap in the $z$ direction. This is modelled by a potential $V_\mathrm{kick}(z,t)=\frac{1}{2}M\alpha z^2\delta(t)$~\cite{Ammann:1997}, with $\alpha$ the kick strength in units of frequency. The approximation $V_\mathrm{kick}(z,t) \propto \delta(t)$ is valid when the kick occurs over a time much less than $\omega^{-1}$, with $\omega$ characterising the time scale of the collective modes of the condensate. The delta kick imparts an inward momentum $M\alpha z$ to the atomic cloud, causing it to contract and become denser. Immediately after the kick, the state is
\begin{equation}\label{eq:kick}
\begin{split}
    \varphi_j(z,\Delta t)&=e^{-\frac{i}{\hbar}\int_0^{\Delta t} V_\mathrm{kick}(z,t)\,dt}\varphi_j(z,0)\\
    &=e^{-\frac{iM}{2\hbar}\alpha z^2}\varphi_j(z,0),
\end{split}
\end{equation}
with $\Delta t\rightarrow 0$ the kick duration. The kicked state provides the initial condition to be used in Eq.~\eqref{eq:timeevolve}, with time evolution obtained by solving Eq.~\eqref{eq:GPEeff} and averaging over truncated Wigner trajectories.

We simulate only the state preparation stage of the full sequence in Fig.~\ref{fig:scheme}(b). This is sufficient for characterising the spin squeezing Eq.~\eqref{xi1} and hence the measurement sensitivity Eq.~\eqref{eq:phase}. The sensitivity calculated this way was shown to agree with that predicted from simulations of the full interferometry sequence as long as $T\gg T_\mathrm{OAT}$~\cite{Szigeti:2020}.

\subsection{Numerical parameters}\label{numerParams}
For concreteness, the two components $j=1,2$ are taken to be hyperfine states $\ket{1}\equiv\ket{F=1,m_F =0}$ and $\ket{2}\equiv\ket{F=2,m_F =0}$ of an $^{87}\mathrm{Rb}$ atom with $(a_{11},a_{22},a_{12})=(100.4,95.0,97.66)a_\mathrm{B}$ ($a_\mathrm{B}$ is the Bohr radius)~\cite{Mertes:2007} and $k_0=1.61\times 10^7$\,m$^{-1}$ ($780$ nm D2 transition)~\cite{Ye:1996}. Equation~\eqref{eq:GPEeff} is solved simultaneously with Eq.~\eqref{eq:Rdynamics} using a fourth-order Runge-Kutta integrator with the kinetic energy term in Eq.~\eqref{eq:GPEeff} evaluated to spectral accuracy. Defining $\varphi_2(z,t)$ in a frame with momentum $\hbar k_0$ [see Eq.~\eqref{eq:psiseparate}] enables both components to be defined on the same momentum grid. We simulate on a periodic grid with $m=4096$ modes, and a system size of $L \sim 500\,\mu\mathrm{m}$, which is large enough to capture both the relative displacement $\sim 2T_\mathrm{OAT}\hbar k_0/M$ and the condensate expansion. Observables are evaluated from $10^4$ truncated Wigner trajectories, which ensures sampling errors are small. We present results for condensates with $N=10^5$ atoms; for appreciably larger condensates, spontaneous scattering can degrade spin squeezing~\cite{Szigeti:2020}. Unless otherwise stated, we set $T_\mathrm{OAT}=10\,\mathrm{ms}$, at which point the condensate has expanded such that OAT interactions are too weak to increase spin squeezing.

\begin{figure}
    \includegraphics[width=\textwidth]{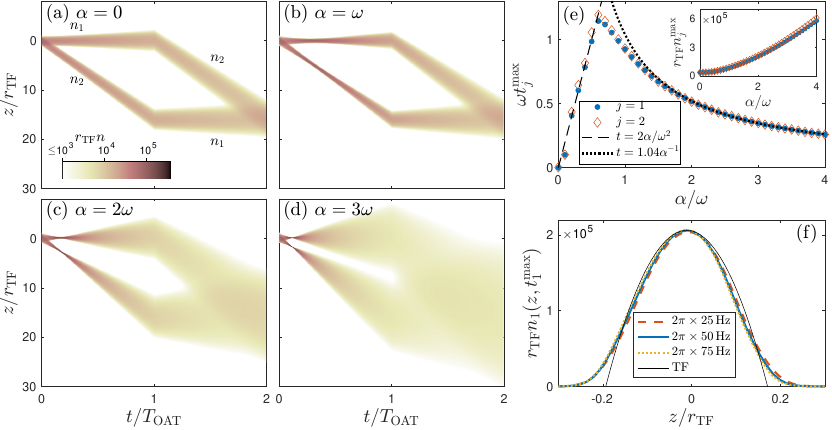}
    \caption{\label{fig:densities} (a)--(d) Condensate density $n(z,t)=n_1(z,t)+n_2(z,t)$ in the free-falling frame for kick strengths $\alpha=0$, $\omega$, $2\omega$ and $3\omega$, respectively. The momentum imparted by the first $\pi/2$ pulse results in the components separating. A $\pi$ pulse is applied at $t=T_\mathrm{OAT}$ and the components recombine at $t\approx 2T_\mathrm{OAT}$. A delta kick increases the peak density of each component. (e) Time $t_j^\mathrm{max}$ (main figure) at which each component reaches its maximum density $n_j^\mathrm{max}$ (inset) as a function of kick strength. Dashed and dotted lines in main figure are fitted scalings $t=2\alpha/\omega^2$ and $t=1.04\alpha^{-1}$, describing $\alpha\ll\omega$ and $\alpha\gtrsim\omega$ respectively. Results in (a)--(e) are for $\omega=2\pi \times 50\,\mathrm{Hz}$. (f) Density profile $n_1(z,t_1^\mathrm{max})$ with a kick strength $\alpha=2\omega$ for three different trap frequencies $\omega=2\pi\times (25,50,75)\,\mathrm{Hz}$. The thin solid line is a scaled Thomas-Fermi profile with matching peak density $n_1^\mathrm{max}$. All results are for $T_\mathrm{OAT}=10\,\mathrm{ms}$ and $N=10^5$, and are obtained from pure GPE simulations.}
\end{figure}

\section{Results}\label{sec:results}

\subsection{Density dynamics during state preparation}\label{sec:density}
Example dynamics for the condensate density during state preparation are shown in Fig.~\ref{fig:densities}(a--d) for varying kick strengths. For simplicity, these results are obtained from simulations of the pure Gross-Pitaevskii equation, Eq.~\eqref{eq:GPEeff} (other numerical parameters as described in Sec.~\ref{numerParams} are unchanged). The resulting density $n(z,t)=n_1(z,t)+n_2(z,t)$, with $n_j(z,t)=|\varphi_j(z,t)|^2$, is very similar to that obtained from full truncated Wigner simulations, since the Wigner noise is small compared to the total number of particles. As expected, the two components initially separate when released from the trapping potential. Then, following a $\pi$ pulse at $t=T_\mathrm{OAT}$, they recombine at $t\approx 2T_\mathrm{OAT}$.

With no delta kick [Fig.~\ref{fig:densities}(a)], the removal of the trapping potential causes the condensate to expand and its density to decrease. As the density decreases, so too do the atom--atom interactions responsible for spin squeezing. A delta kick, by contrast, causes the condensate to initially contract, resulting in a focussing of each component that depends on the kick strength $\alpha$ [Fig.~\ref{fig:densities}(b--d)]. Figure~\ref{fig:densities}(e) shows the time $t_j^\mathrm{max}$ at which $n_j(z,t)$ is a maximum $n_j^\mathrm{max}$, and exhibits two qualitatively distinct regimes. For $\alpha\ll \omega$, the initial inward velocity $v_0=\alpha z$ imparted by the kick is slowed by the initial outward acceleration $a_0=\omega^2 z$ from interactions. The deceleration occurs in a time $v_0/a_0$, hence the focussing time scales as $\sim \alpha/\omega^2$. For $\alpha\gtrsim \omega$ the cloud velocity is dominated by the kick and the focussing time scales as $\sim z/v_0=\alpha^{-1}$~\cite{Dupays:2021}. The maximum density at the focus, $n_j^\mathrm{max}$, shown in the inset to Fig.~\ref{fig:densities}(e), increases with $\alpha$.

For a fixed $\alpha/\omega$, the density profile at the focus shows little sensitivity to the initial trap frequency after scaling by $r_\mathrm{TF}$, see Fig.~\ref{fig:densities}(f). This is consistent with a scaling solution
\begin{equation}\label{eq:nscaling}
    n_j(z,t)\sim \frac{1}{r_\mathrm{TF}}\zeta_j\left(\frac{z}{r_\mathrm{TF}},\omega t\right),
\end{equation}
with $\zeta_j$ dependent on $\alpha/\omega$. Note, however, that the density deviates slightly from the typical scaled Thomas-Fermi ansatz~\cite{Castin:1996}, see Fig.~\ref{fig:densities}(f), indicating the presence of collective excitations~\cite{Stringari:1996}.

\begin{figure}
    \includegraphics[width=\textwidth]{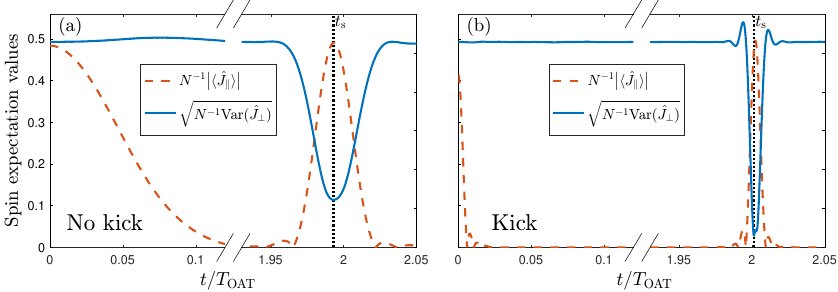}
    \caption{\label{fig:spins} Average spin $N^{-1}\big|\langle\hat{J}_\parallel(t)\rangle\big|$ (dashed line) and standard deviation $\sqrt{N^{-1}\operatorname{Var}(\hat{J}_\perp(t))}$ (solid line) for (a) no delta kick and (b) a delta kick of strength $\alpha=2\omega$. The ratio of the quantities plotted gives the spin squeezing Eq.~\eqref{xi1}. The squeezing is optimal ($\xi$ is a minimum) at a time $t_\mathrm{s}\approx 2T_\mathrm{OAT}$ (vertical dotted line). The increased density due to the delta kick reduces the variance $\operatorname{Var}(\hat{J}_\perp(t_\mathrm{s}))$ with no appreciable affect on $\big|\langle\hat{J}_\parallel(t_\mathrm{s})\rangle\big|$, resulting in enhanced spin squeezing. All results are for $\omega=2\pi \times 50\,\mathrm{Hz}$, $T_\mathrm{OAT}=10\,\mathrm{ms}$ and $N=10^5$. Note the break in the horizontal axis.}
\end{figure}

\subsection{Delta-kick spin squeezing}\label{sec:squeezing}
Quantifying the spin squeezing accumulated during state preparation requires simulations beyond pure GPE; here, we use the truncated Wigner method (see Sec.~\ref{sec:TW}). The dynamics of spin expectation values $N^{-1}\big|\langle\hat{J}_\parallel(t)\rangle\big|$ and $\sqrt{N^{-1}\operatorname{Var}(\hat{J}_\perp(t))}$ from truncated Wigner simulations are shown in Fig.~\ref{fig:spins} for both no kick [(a)] and a kick of strength $\alpha=2\omega$ [(b)]. The ratio of the quantities plotted gives the spin squeezing [Eq.~\eqref{xi1}]. The initial spin coherent state has $\langle\hat{J}_\parallel\rangle\approx\langle\hat{J}_x\rangle=N/2$ and $\operatorname{Var}(\hat{J}_\perp)=N/4$. As the components separate, $\big|\langle\hat{J}_\parallel\rangle\big|$ rapidly decreases due to the absence of component overlap. When the two components recombine at $t\approx 2T_\mathrm{OAT}$, $\big|\langle\hat{J}_\parallel\rangle\big|$ returns to close to its initial value, while the variance $\operatorname{Var}(\hat{J}_\perp)$ is reduced due to the accumulated OAT. The spin distribution is therefore squeezed ($\xi<1$). The minimum spin variance $\operatorname{Var}(\hat{J}_\perp)$ in Fig.~\ref{fig:spins}(b) is $\approx 8\times 10^{-4}N$ --- a 25 dB reduction from a spin coherent state and a factor of 16 smaller than the variance with no kick. Optimal squeezing occurs at a time $t_\mathrm{s}$, shown in Fig.~\ref{fig:spins}, which is ideally as close as possible to $2T_\mathrm{OAT}$ to avoid the need for careful calibration. The delta kick results in a reduced atomic density when the two components recombine [see Fig.~\ref{fig:densities}(a-d)], and hence $t_\mathrm{s}$ is closer to $2T_\mathrm{OAT}$.

\begin{figure}
    \includegraphics[width=\textwidth]{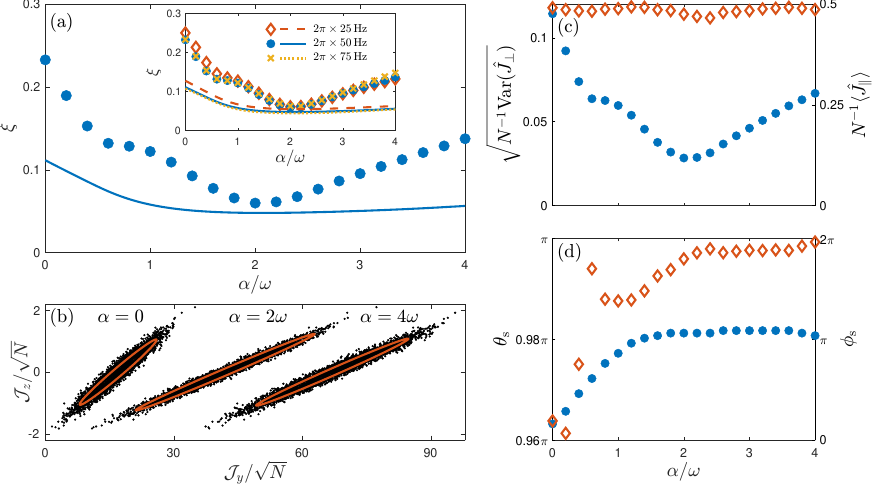}
    \caption{\label{fig:kick} (a) Main figure: Spin-squeezing parameter [Eq.~\eqref{xi1}] as a function of kick strength $\alpha$ obtained from truncated Wigner simulations (circles) with $\omega=2\pi\times 50\,\mathrm{Hz}$. The squeezing is optimal ($\xi$ is smallest) for a kick strength $\alpha\approx 2\omega$. The solid line is the two-mode prediction Eq.~\eqref{eq:spsq}, with $\lambda$ and $Q$ obtained from pure GPE simulations. Inset: As in main figure, but for initial trap frequencies $\omega=2\pi\times (25,50,75)\,\mathrm{Hz}$ [symbols are truncated Wigner simulations, lines are Eq.~\eqref{eq:spsq}]. (b) Scatter plot of $(\mathcal{J}_y,\mathcal{J}_z)$ at $t=t_\mathrm{s}$ for $\alpha=0$ (left), $2\omega$ (middle) and $4\omega$ (right), obtained from $10^4$ truncated Wigner trajectories and offset horizontally for clarity. For simplicity we remove equatorial drift by aligning $\hat{J}_\parallel$ with $\hat{J}_x$. The red ellipse is computed from the two-mode approximation and has semi-minor and semi-major axes of $\xi+m/8$ and $\xi^{-1}+m/8$ respectively, with $m/8$ accounting for normal ordering [see Eq.~\eqref{eq:TWexpect}]. (c) Spin expectation values $\sqrt{N^{-1}\operatorname{Var}(\hat{J}_\perp(t_\mathrm{s}))}$ (filled circles, left axis) and  $N^{-1}|\langle \hat{J}_\parallel(t_\mathrm{s})\rangle|$ (empty diamonds, right axis) for varying kick strength. (d) Corresponding angles $\theta_\mathrm{s}$ (filled circles, left axis) and $\phi_\mathrm{s}$ (empty diamonds, right axis). Results in (b-d) are for $\omega=2\pi \times 50\,\mathrm{Hz}$.  All results are for $T_\mathrm{OAT}=10\,\mathrm{ms}$ and $N=10^5$.}
\end{figure}

The spin squeezing as a function of kick strength is quantified more precisely in Fig.~\ref{fig:kick}(a), which shows the squeezing parameter [Eq.~\eqref{xi1}] obtained from truncated Wigner simulations (filled circles). Optimal spin squeezing is achieved for a kick strength $\alpha\approx 2\omega$, at which point $\xi\approx 0.06$ is four times smaller than the scheme with no kick ($\alpha=0$). This corresponds to a phase sensitivity a factor $1/\xi\approx 17$ below the standard quantum limit [see Eq.~\eqref{eq:phase}]. Squeezing gradually degrades for stronger kick strengths $\alpha>2\omega$, which will be discussed in Sec.~\ref{sec:singlemode}. The optimal squeezing at $\alpha\approx 2\omega$ is also evident in scatter plots of $\mathcal{J}_\mu$, which are shown in Fig.~\ref{fig:kick}(b) for $\alpha=0$, $2\omega$ and $4\omega$.

For completeness, the corresponding variance $\operatorname{Var}(\hat{J}_\perp(t_\mathrm{s}))$ and spin $|\langle\hat{J}_\parallel(t_\mathrm{s})\rangle|$ are shown in Fig.~\ref{fig:kick}(c). The degradation in squeezing for $\alpha>2\omega$ can be traced to an increase in $\operatorname{Var}(\hat{J}_\perp)$ (as opposed to a decrease in $\big|\langle\hat{J}_\parallel\rangle\big|$). The angles $\theta_\mathrm{s}$ and $\phi_\mathrm{s}$, which describe the orientation of the spin distribution, are shown in Fig.~\ref{fig:kick}(d). The squeezing occurs in a spin direction off-axis from $\hat{J}_z$ by a small angle $\pi-\theta_\mathrm{s}\ll 1$ due to shearing under OAT [see Fig.~\ref{fig:kick}(b)], which plateaus at $\theta_\mathrm{s}\approx 0.98 \pi$ for $\alpha\gtrsim 1.6\omega$. The dependence of $\phi_\mathrm{s}$ on kick strength is primarily due to the phase shift $(2T_\mathrm{OAT}-t_\mathrm{s})\hbar k_0^2/(2M)$, which arises when the state preparation is not perfectly symmetric around the mirror M1 ($t_\mathrm{s}\ne 2T_\mathrm{OAT})$.

Very similar spin squeezing is obtained for different initial trap frequencies when $\alpha$ is scaled by $\omega$, see inset to Fig.~\ref{fig:kick}(a). This is consistent with dynamics predominantly in the hydrodynamic regime, in which case the condensate evolves with a characteristic timescale $\omega^{-1}$~\cite{Stringari:1996} and squeezing is independent of $\omega$ as long as $T_\mathrm{OAT}\gg\omega^{-1}$.

\subsection{Comparison with two-mode dynamics}\label{sec:singlemode}
To gain additional insight, we compare the truncated Wigner simulations to a two-mode approximation~\cite{Haine:2014,Szigeti:2020,Haine:2005b}, obtained by setting
\begin{subequations}
\begin{align}
    \hat{\psi}_1(\mathbf{r},t)&=u_1(\mathbf{r},t)\hat{a}_1+\hat{v}_1(\mathbf{r},t),\\
    \hat{\psi}_2(\mathbf{r},t)&=u_2(\mathbf{r},t)e^{ik_0z}\hat{a}_2+\hat{v}_2(\mathbf{r},t).
\end{align}
\end{subequations}
Here $\hat{a}_j$ are bosonic lowering operators for mode $\ket{j}$, with mode function $u_j(\mathbf{r},t)$ satisfying $\int|u_j(\mathbf{r},t)|^2\,d\mathbf{r}=1$, and $\hat{v}_j(\mathbf{r},t)$ are vacuum field operators with commutation relations $[\hat{v}_j(\mathbf{r},t),\hat{v}_\ell^\dagger(\mathbf{r}',t)]=\delta_{j\ell}(\delta(\mathbf{r}-\mathbf{r}^\prime)-u_j(\mathbf{r})u_\ell(\mathbf{r}'))$. The state-preparation dynamics in the two-mode regime corresponds to an OAT Hamiltonian~\cite{Sorensen:2001b,Kitagawa:1993}
\begin{equation}\label{eq:Hoat}
    \hat{H} = \hbar \chi(t) \hat{J}^2_z
    \end{equation}
with
\begin{equation}\label{eq:chidef}
    \chi(t)=\chi_{11}(t) +\chi_{22}(t) - 2\chi_{12}(t)
\end{equation}
the squeezing rate dependent on both intra- and inter-component terms $\chi_{j\ell}(t) = \frac{g_{j\ell}}{2\hbar}\int |u_j(\textbf{r},t)|^2 |u_\ell(\textbf{r},t)|^2\,d\textbf{r}$. Note $\chi$ is largest when $\chi_{11}$ and $\chi_{22}$ are large but $\chi_{12}$ is small, i.e.\ the two components are spatially separated. A delta kick enables the peak density to occur when the two components are spatially separated, whereas with no kick the peak density occurs when the two components overlap [see Fig.~\ref{fig:densities}(a-d)].

Evolving the initial state $2^{-N/2}(\ket{1}+ \ket{2})^{\bigotimes N}$ with Eq.~\eqref{eq:Hoat} and evaluating the optimal spin squeezing, Eq.~\eqref{xi1}, gives~\cite{Szigeti:2020}
\begin{eqnarray}\label{eq:spsq}
\xi \approx \frac{\sqrt{1-\frac{1}{2}|Q|N\lambda(\sqrt{4+|Q|^2N^2\lambda^2} -|Q|N\lambda)}}{|Q|},
\end{eqnarray}
where
\begin{equation}\label{eq:lamdef}
    \lambda=\int_0^{2T_\mathrm{OAT}}\chi(t)\,dt
\end{equation}
is the OAT factor, and
\begin{equation}\label{eq:Qdef}
    Q = \int u^{*}_1(\textbf{r},2\mathrm{T_{OAT}}) u_2(\textbf{r},2\mathrm{T_{OAT}})\,d\textbf{r} 
\end{equation}
is the final overlap. The squeezing, Eq.~\eqref{eq:spsq}, increases monotonically with the OAT factor $\lambda$.

We compare the truncated Wigner results with Eq.~\eqref{eq:spsq} in Fig.~\ref{fig:kick}(a). The modes $u_j(\mathbf{r},t)$ used to compute $\lambda$ [Eq.~\eqref{eq:lamdef}] and $Q$ [Eq.~\eqref{eq:Qdef}] are obtained from pure GPE simulations [Eq.~\eqref{eq:GPEeff}], identifying $u_j(\mathbf{r},t)=\psi_\perp(\rho,t)\varphi_j(z,t)/\sqrt{N_j}$, with $N_j=\int|\psi_\perp(\rho,t)|^2|\varphi_j(z,t)|^2\,d\textbf{r}$~\cite{Szigeti:2020}. Typically, the two-mode prediction Eq.~\eqref{eq:spsq} overestimates the degree of spin squeezing, except at the optimal kick strength $\alpha\approx 2\omega$ for which the two-mode and truncated Wigner results are quantitatively similar. An estimate of the spin distribution for the two-mode approximation is included in Fig.~\ref{fig:kick}(b), and describes well the truncated Wigner distribution for $\alpha=2\omega$ but overestimates squeezing otherwise. Like the truncated Wigner simulations, the two-mode prediction Eq.~\eqref{eq:spsq} shows little sensitivity to initial trap frequency [see Fig.~\ref{fig:kick}(a) inset], which follows directly from the scaling solution Eq.~\eqref{eq:nscaling}.\footnote{Explicitly, Eq.~\eqref{eq:nscaling} gives $\int_0^{2T_\mathrm{OAT}}\chi_{j\ell}(t)\,dt\sim\frac{2M\omega}{3\pi\mu}\int_0^\infty f_{j\ell}(u)\,du$ with $f_{j\ell}(\omega t)=\frac{g_{j\ell}}{N_jN_\ell b(\omega t)^2}\int \zeta_j(u,\omega t)\zeta_\ell(u,\omega t)\,du$, and we assume that $f(u)\approx 0$ for $u>2\omega T_\mathrm{OAT}$. Assuming a Thomas-Fermi profile for the initial density $n(z,0)$ gives $\mu\propto\omega$, in which case $\lambda$ depends only on trap frequency through $\alpha/\omega$.}

\begin{figure}
    \includegraphics[width=\textwidth]{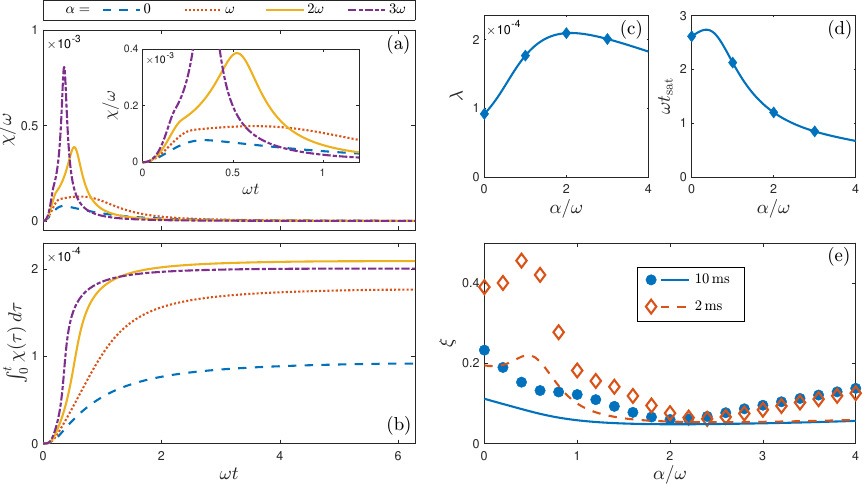}
    \caption{\label{fig:gpe} (a) Squeezing rate Eq.~\eqref{eq:chidef} for kick strengths $\alpha=0,\,\omega,\,2\omega,\,3\omega$ estimated from pure GPE simulations. The delta kick increases the peak squeezing rate. For $\alpha\gtrsim 2\omega$, the delta kick also reduces the duration for which the squeezing rate is appreciable. Inset: Zoomed region $\omega t\le 1.2$. (b) Corresponding integral $\int_0^t\chi(\tau)\,d\tau$, which gives the OAT factor $\lambda$ at $t=2T_\mathrm{OAT}$ [Eq.~\eqref{eq:lamdef}]. (c) The OAT factor peaks at $\alpha\approx 2\omega$, coinciding with the optimal spin squeezing in Fig.~\ref{fig:kick}(a). (d) The saturation time $t_\mathrm{sat}$, defined via $\int_0^{t_\mathrm{sat}}\chi(\tau)\,d\tau=0.9\lambda$, decreases with increasing kick strength for $\alpha\gtrsim 0.4\omega$. Diamonds in (c) and (d) correspond to the kick strengths presented in (a,b). Results in (a--d) are for $T_\mathrm{OAT}=10\,\mathrm{ms}$. (e) Spin-squeezing parameter as a function of kick strength for $T_\mathrm{OAT}=10\,\operatorname{ms}\approx 3\omega^{-1}$ (filled circles) and $T_\mathrm{OAT}=2\,\operatorname{ms}\approx 0.6\omega^{-1}$ (empty diamonds). With no kick, squeezing is reduced by approximately half for $T_\mathrm{OAT}=2\,\operatorname{ms}$, compared to $T_\mathrm{OAT}=10\,\mathrm{ms}$. For kick strengths $\alpha\gtrsim 2\omega$, squeezing is similar for both $T_\mathrm{OAT}$ values. The two-mode prediction Eq.~\eqref{eq:spsq} for $T_\mathrm{OAT}=10\,\operatorname{ms}$ (solid line) and $T_\mathrm{OAT}=2\,\operatorname{ms}$ (dashed line) is also shown. All results are for $\omega=2\pi \times 50\,\mathrm{Hz}$ and $N=10^5$.}
\end{figure}

Spin squeezing degrades for increasing kick strength $\alpha>2\omega$ for both the truncated Wigner results and the two-mode approximation, see Fig.~\ref{fig:kick}(a). The small degradation in the two-mode approximation can be explained by considering both the magnitude and duration of the squeezing rate, shown in Fig.~\ref{fig:gpe}(a), which when integrated over the state preparation [see Fig.~\ref{fig:gpe}(b)] gives $\lambda$ [Eq.~\eqref{eq:lamdef}]. Although the higher peak density from a delta kick results in a higher maximum squeezing rate, the subsequent expansion is also accelerated [see Fig.~\ref{fig:densities}(a--d)], resulting in reduced duration of OAT interactions. The combination of these two effects results in a peak in $\lambda$ at a kick strength $\alpha\approx 2\omega$, see Fig.~\ref{fig:gpe}(c), which coincides with the optimal spin squeezing in Fig.~\ref{fig:kick}(a). The degradation of squeezing is more appreciable in the truncated Wigner results. For $\alpha>2\omega$, the focussing time $\approx\alpha^{-1}$ [Fig.~\ref{fig:densities}(e)] is smaller than the time scale $\omega^{-1}$ of hydrodynamic collective modes, which likely results in significant excitation of the collective modes and deviation from the two-mode approximation.

Finally, we discuss how a delta kick enables the state preparation time $2T_\mathrm{OAT}$ to be reduced without compromising squeezing. For $2T_\mathrm{OAT}\gg\omega^{-1}$, the integral of $\chi(t)$ increases up to a saturation time $t_\mathrm{sat}$, beyond which the atomic cloud is too dilute to give substantial OAT. Figure~\ref{fig:gpe}(b) shows that the saturation time tends to decrease with increasing kick strength, consistent with the increased expansion rate following a kick [Fig.~\ref{fig:densities}(a--d)]. To quantify this, we define $t_\mathrm{sat}$ via $\int_0^{t_\mathrm{sat}}\chi(\tau)\,d\tau=0.9\lambda$, which is approximately independent of $T_\mathrm{OAT}$ as long as $2T_\mathrm{OAT}\gg\omega^{-1}$, and decreases with $\alpha$ for $\alpha\gtrsim 0.4\omega$, see Fig.~\ref{fig:gpe}(d). A smaller $t_\mathrm{sat}$ is advantageous, as spin squeezing accumulates more rapidly and the state preparation time can be reduced. This is demonstrated in Fig.~\ref{fig:gpe}(e), which shows the spin-squeezing parameter as a function of kick strength for $T_\mathrm{OAT}=10\,\operatorname{ms}\approx 3\omega^{-1}$ and $T_\mathrm{OAT}=2\,\operatorname{ms}\approx 0.6\omega^{-1}$, in both the truncated Wigner and two-mode approximation. With no kick, spin squeezing is reduced by approximately half with the smaller $T_\mathrm{OAT}$, as there is insufficient time for OAT. For a kick strength $\alpha\gtrsim 2\omega$, the OAT occurs more quickly and both $T_\mathrm{OAT}$ values give similar squeezing. Note that a kick can degrade squeezing if the focussing increases the density at the second beamsplitter BS2, which reduces component overlap. For $T_\mathrm{OAT}=2\,\operatorname{ms}\approx 0.6\omega^{-1}$, this occurs for $\alpha\lesssim 0.6\omega$ [see Fig.~\ref{fig:densities}(e)], resulting in the degradation in squeezing for small kick strengths in Fig.~\ref{fig:gpe}(e).

\section{Conclusion}\label{sec:conclusion}
We have presented an improved cold-atom gravimetry scheme, which focusses the condensate with a delta kick to enhance spin squeezing and therefore phase sensitivity. We quantify the squeezing using multi-mode truncated Wigner simulations, obtaining a factor of $\sim 20$ improvement in phase sensitivity compared to the standard quantum limit. At the optimal kick strength, spin squeezing is enhanced fourfold compared to no kick, and is close to that obtained from a two-mode approximation. A delta-kick also reduces the duration of OAT interactions, allowing for the state preparation time $2T_\mathrm{OAT}$ to be reduced and the optimal squeezing time $t_\mathrm{s}$ to be more easily calibrated.

The truncated Wigner results presented here neglect scattering beyond one dimension, which may degrade spin squeezing~\cite{Szigeti:2020}. Four-wave mixing, for example, may give rise to halo scattering when the two components overlap~\cite{Chikkatur:2000,Norrie:2005}. An advantage of the delta-kick scheme is that the density of the initial overlapping clouds can be low, reducing halo scattering. We have compared multi-mode truncated Wigner simulations with a two-mode approximation. A more accurate approximation may be possible by considering time-dependent Bogoliubov excitations around the two-mode regime, which would also enable exploration of multi-mode effects in three dimensions. The scale invariance of our results with trap frequency suggests that the relevant excitations will be hydrodynamic modes~\cite{Stringari:1996}.

\section*{Acknowledgements}
We acknowledge useful discussions with Stuart Szigeti, Zain Mehdi, and Joe Hope.

\paragraph{Funding information}
This research was supported by the Australian government Department of Industry, Science, and Resources via the Australia-India Strategic Research Fund (AIRXIV000025) and the Australian Research Council Centre of Excellence for Engineered Quantum Systems (project ID CE170100009). S.A.H. is supported through Australian Research Council Future Fellowship, Grant No. FT210100809.

\end{document}